\documentclass[aip,reprint
,amsmath,amssymb
]{revtex4-1}

\usepackage{graphicx}% Include figure files
\usepackage{xcolor}
\usepackage{dcolumn}% Align table columns on decimal point
\usepackage{bm}% bold math
\usepackage[utf8]{inputenc}
\usepackage[T1]{fontenc}
\usepackage{mathptmx}
\usepackage{etoolbox,hyperref}

\draft % marks overfull lines with a black rule on the right

\begin{document}

%\preprint{}

\title[Generalized nonlinear q-voter model]{Modeling biases in binary decision-making within the generalized nonlinear q-voter model} %Title of paper



\author{Maciej Doniec}
\affiliation{Department of Science, Technology and Society Studies, Wrocław University of Science and Technology, Poland}

\author{Pratik Mullick}
\email{pratik.mullick@pwr.edu.pl, corresponding author}
\affiliation{Department of Operations Research and Business Intelligence, Wrocław University of Science and Technology, Poland}%Department and Organization
%            addressline={ul. I. Łukasiewicza 3/5}, 
%            city={Wrocław},
%            postcode={50-371}, 
%            state={Lower Silesia},
%            country={Poland}}

\author{Parongama Sen}
\affiliation{Department of Physics, University of Calcutta, India}%Department and Organization
%             addressline={92 Acharya Prafulla Chandra Road}, 
%             city={Kolkata},
%             postcode={700009}, 
%             state={West Bengal},
%             country={India}}

\author{Katarzyna Sznajd-Weron}
\affiliation{Department of Science, Technology and Society Studies, Wrocław University of Science and Technology, Poland}


\date{\today}

\begin{abstract}
Collective decision-making is a process by which a group of individuals determines a shared outcome that shapes societal dynamics; from innovation diffusion to organizational choices. A common approach to model these processes is using binary dynamics, where the choices are reduced to two alternatives. One of the most popular models in this context is the $q$-voter model, which assumes that opinion changes are driven by peer pressure from a unanimous group. However, real-world decisions are also shaped by prior personal choices and external influences, such as mass media, which introduce biases that can favor certain options over others. To address this, we propose a generalized $q$-voter model that incorporates these biases. In our model, when the influence group is not unanimous, the probability that an individual changes its opinion depends on its current state, breaking the symmetry between opinions. In limiting cases, our model recovers both the original $q$-voter model and several recently introduced modifications of the $q$-voter model, while extending the framework to capture a broader range of scenarios. We analyze the model on a complete graph using analytical methods and Monte Carlo simulations. Our results highlight two key findings: (1) for larger influence groups ($q>3$), a phase emerges where both adopted and partially adopted states coexist, (2) in small systems, greater initial support for an opinion does not necessarily increase its likelihood of widespread adoption, as reflected in the unique form of the exit probability. These results point to one of the key issues in social science, the importance of group size in collective action.

\end{abstract}


\maketitle %\maketitle must follow title, authors, abstract and \pacs

 \begin{quotation}
% The ``lead paragraph'' is encapsulated with the \LaTeX\ 
% \verb+quotation+ environment and is formatted as a single paragraph before the first section heading. 
% (The \verb+quotation+ environment reverts to its usual meaning after the first sectioning command.) 
% Note that numbered references are allowed in the lead paragraph.
% %
% The lead paragraph will only be found in an article being prepared for the journal \textit{Chaos}.

Collective decision-making is a fundamental process in social systems, from consumer choices to political elections, where the opinions of individuals shape the final outcome. A well-established approach to modeling such dynamics is the $q$-voter model, which describes how people adopt opinions based on group influence. However, in reality, individuals are often influenced not only by their peers but also by their own previous choices or external pressures, such as advertising or political campaigns. In this work, we propose a generalized $q$-voter model that accounts for this self-reinforcing tendency when individuals face conflicting opinions. We find that this internal bias can prevent full consensus and instead lead to stable coexistence of competing opinions. Moreover, in small groups, increasing initial support for an opinion does not always ensure its eventual dominance, revealing that both group size and the strength of individual bias can fundamentally shape collective outcomes. These results contribute to a better understanding of how personal preferences and group influence together drive opinion formation in social and organizational settings.

 \end{quotation}

%\section{}
%\label{}
%\subsection{}
%\subsubsection{}

\section{Introduction}\label{intro}
Binary decision-making is a fundamental method that simplifies choices into two clear alternatives, facilitating faster and more efficient decision processes. Psychologically, this approach helps mitigate decision fatigue, the phenomenon in which decision quality declines after making numerous choices \cite{Hir:etal:19}. Cognitive load theory supports this approach, suggesting that memory has limited capacity. When overwhelmed with excessive information or complex tasks, cognitive efficiency decreases significantly \cite{Sweller1988}. Thus, narrowing options to a binary format reduces extraneous cognitive load, allowing people to concentrate on fundamental aspects without distraction from irrelevant details. Recent studies emphasize this psychological benefit \cite{Ball:23}. The binary response format is also very important from the perspective of statistical physics and, by extension, sociophysics (called also social physics) — a framework that integrates methods from statistical physics with the modeling of complex social phenomena \cite{JUSUP20221,CHICA2024109254}.

% \rev{From a broader perspective, social physics has emerged as a framework integrating statistical physics methods to model complex societal behaviors, including decision-making processes\cite{JUSUP20221}.}

% Furthermore, recent studies highlight how opinion formation, when coupled with strategic decision-making in social dilemmas, can drive social tensions and polarization, showing the interplay between decision dynamics and social influence \cite{CHICA2024109254}.

In many cases, a decision-making is a sequential process that involves a series of decisions made over time. This has important consequences, because psychological experiments show that our current perceptions and decisions are shaped by recent experiences, which is known as sequential bias \cite{Che:Che:Shi:24}. For example, when items are presented in a sequence, the evaluation of each item is influenced by the value of the previous item and the response given before it \cite{Kra:Pus:21}. Another source of bias in sequential decision making is related to social influence and is therefore particularly relevant in collective decision contexts. The results of a laboratory experiment show that when people say which of two answers they think is correct, the majority of them are more likely to be wrong if they can see how often those two answers have been chosen by previous people than if they have no such opportunity \cite{Frey2021}.

Therefore, to realistically describe the process of opinion formation or decision-making, we need to account for various biases related both to individual past experiences and responses, as well as to the responses of others. Potentially, we could also incorporate additional strategies, as was recently done for the continuous opinion dynamics model with bounded confidence, in which opinion formation was coupled with strategy selection in social dilemmas \cite{CHICA2024109254}. However, in this paper, we focus on non-strategic opinion dynamics \cite{Gra:Rus:20}. We are aware that, from a psychological point of view, opinion formation and decision making are two different processes. However, we will use the two terms interchangeably here, much as is often done for \textit{opinions}, \textit{beliefs}, and \textit{attitudes}. In fact, the models used to study all of them often overlap \cite{Ols:Gal:24}.

Among several nonstrategic models of binary opinion dynamics \cite{Gleeson2013,jed:szn:19}, the $q$-voter model \cite{qvm_castellano}, also known as the nonlinear voter model \cite{Peralta2018,ramirez2024ordering}, is one of the most universal, and several biases have already been considered within this framework, as will be reviewed in the next section. However, to our knowledge, the $q$-voter model \cite{qvm_castellano} has not been thoroughly studied under the influence of a sequential bias. This inspired us to generalize the model to incorporate such an effect, which not only breaks the symmetry between two alternative choices but also accounts for cases where the evaluation of a current choice is influenced by previous decisions. 

The remainder of the paper is structured as follows. In the next section, we briefly review the literature on linear and non-linear voter models, particularly focusing on how various biases have been modeled, to place our research in a broader context. We then introduce the new generalized model, which, in special cases, reduces to the original $q$-voter model \cite{qvm_castellano} and the recently proposed biased version with influence from the media \cite{MUSLIM2024129358}. Next, we derive non-linear differential equations that describe the aggregated behavior of the model, identify fixed points, and analyze their stability. This enables us to construct the phase diagram, which reveals asymmetrical phases not present in the original $q$-voter model. Finally, we derive the exit probability for finite systems and demonstrate that for $q \ge 3$, it takes on a unique form that has not been observed in previous models. We conclude with a discussion of the theoretical and practical applications of our model.

\section{Related literature}\label{lit.rev}
The foundational voter model \cite{Holley1975ergodic,liggett1985interacting} involves binary opinions in which each agent randomly adopts the opinion of a neighboring agent. This model exhibits a linear behavior in exit probabilities, as the probability of consensus is directly proportional to the initial fraction of agents holding a particular opinion. The linear voter model has been extensively studied due to its simplicity and ability to capture fundamental aspects of opinion dynamics in social systems \cite{Redner2019275}. Variations include the introduction of ``zealots" who never change their opinions, influencing the overall consensus time. \cite{mobilia2003does} Another variant in literature is the biased voter model \cite{ferreira1990probability,czaplicka2022biased,durocher2022invasion}, where agents favor one opinion over another. For example, Czaplicka et al. \cite{czaplicka2022biased} divide agents into fixed biased and unbiased groups: unbiased agents follow the original voter dynamics, while biased agents flip their opinions with probabilities that depend on their current opinion and that of their neighbor. 

Partisan voter models \cite{masuda2010heterogeneous,llabres2023partisan} extend the original voter model by introducing agents with inherent biases or preferences towards specific opinions.
The noisy voter model \cite{GRANOVSKY1995}, allowing spontaneous opinion flips, represents yet another adaptation of the voter framework. Linear voter models have also been analyzed on diverse network structures, such as lattices, random graphs, and scale-free networks, revealing how topology influences convergence to consensus \cite{Sood2005,sood2008voter}.

Castellano et al. \cite{qvm_castellano} proposed a nonlinear extension of the voter model, called the $q$-voter model, where an agent interacts with a selected group of $q$ neighbors. If all $q$ neighbors share the same opinion, the agent conforms by adopting that opinion; otherwise, the agent’s opinion flips with a set probability. This model is different from the majority rule model \cite{galam2008sociophysics,krapivsky2021divergence}, where a $q$-panel adopts the majority opinion within the panel.

A large number of studies have examined variants of the nonlinear $q$-voter model, adapting its original formulation to different contexts. For example, Nyczka et al.\cite{nyczka2012phase} examined three different versions of the model that included two types of social responses that occur with complementary probabilities: conformity and one type of nonconformity (two types of anticonformity and independence). Timpanaro and Prado \cite{timpanaro2014exit} explored a conformist $q$-voter model on a one-dimensional lattice with periodic boundary conditions, randomly selecting a successive $q$-panel, where unanimous opinions influence neighboring agents or a single neighboring agent. Javarone et. al\cite{javarone2015conformism} studied a variation of the $q$-voter model in which agents are conformists with some probability and anticonformists with a complementary probability, without introducing independent agents. On the other hand, Mobilia \cite{mobilia2015nonlinear} studied the $q$-voter model on a complete graph with ``zealots'' representing independent agents who are not susceptible to social influence.  Byrka et al. \cite{byrka2016difficulty} proposed a modified model in which each agent is subject to social influence or to some external factor (ignoring its $q$ neighbors) with complementary probabilities.  Recently, another approach to nonconformism has been proposed by introducing a skepticism parameter that quantifies an agent's propensity to be a nonconformist \cite{Anu:etal:25}. This parameter influences how likely an agent is to resist social influence, affecting both anticonformity (going against the majority) and independence (randomly choosing an opinion). 

The $q$-voter model and its variants have also been studied in multiplex networks \cite{gradowski2020pair}, duplex cliques \cite{chmiel2015phase}, random graphs \cite{vieira2020pair,Krawiecki2024}, Barabasi-Albert networks \cite{Fardela2024} and scale-free networks \cite{vieira2020pair,Anu:etal:25}. Recently, the $q$-voter model was also used to identify an effective seeding strategy  on social networks \cite{Lip:24}.

A notable variant, the threshold $q$-voter model, examined by Vieira et al. \cite{vieira2018threshold}, was tested on a complete graph; in this model, unanimity among a minimum threshold $q_0$ of $q$ agents ($0\leq q_0 \leq q$) is enough to sway the focal agent, while also allowing the agent to remain independent. This threshold model was later applied to random networks \cite{vieira2020pair}. Another extension by Nyczka et al. \cite{nyczka2018conformity} presented a generalized threshold $q$-voter model that incorporated basic social responses such as conformity, anticonformity, independence, and uniformity/congruence. A recent work \cite{vasconcelos2019consensus} introduced a model to explore how opinion dynamics is affected by complex contagion, where individuals adopt opinions based on reinforcement from neighboring contacts. This model extends the voter model framework by incorporating complex contagion processes that require a threshold level of social support for an opinion to spread.

Some recent studies on multistate voter models extend the classic binary voter model to scenarios where agents can adopt one of several discrete states or opinions \cite{now:szn:22}. Models where agents can hold more than two opinions have been studied to explore how the number of available states impacts consensus, coexistence, or fragmentation. Ramirez et al. \cite{ramirez2022local} investigated multistate voter models on complete graphs and networks, analyzing the dynamics of consensus formation and the coarsening process through a sequence of metastable states. In a related study, Ramirez et al. \cite{ramirez2024ordering} examined how nonlinear extensions of multistate voter models impact consensus dynamics, highlighting differences in ordering behavior compared to linear models.

Of late, $q$-voter models \cite{MUSLIM2024129358,Fardela2024} were analyzed, where if the $q$-panel is not unanimous, the focal agent could adopt an opinion influenced by mass media with a certain probability. In another variant \cite{mullick2025social}, in case of lack of unanimity in the $q$-panel, the focal agent chooses an opinion determined by the weighted influence of agents in the $q$-panel. A related model studied by Civitarese \cite{civitarese2021external} explores how an independent agent might question its own stance when affected by an unreliable external influence, similar to societal effects of mass media on decision-making. Galam and Cheon in their asymmetric contrarian model \cite{galam2020asymmetric} introduce a fraction of agents, who oppose the local majority in group discussions. They introduce asymmetry by allowing different proportions of contrarians in the two opinion camps. This asymmetry in contrarian behavior plays a role analogous to the bias in our generalized $q$-voter model. On the other hand, in the contrarian majority rule model \cite{gimenez2023contrarian} by Gimenez et. al., asymmetry enters through an external oscillating propaganda field rather than a fixed internal bias.

\section{Description of the model}\label{model}
%In 2009 Castellano et al. introduced a $q$-voter model, being a natural generalization of the linear voter model \citep{qvm_castellano}. This was already written in the previous section
We consider a population of $N$ agents on a complete graph, where each agent is described by a dynamic binary variable $S_i=\pm 1 (\uparrow/\downarrow), \;i=1,\ldots,N$ representing a binary opinion (yes/no, agree/disagree), a decision such as adopting/not adopting a certain innovation (technology, practice, etc.), or just a state adopted/not adopted (unadopted). In this paper we will use the latter nomenclature.  The state of an agent -- referred to here as the target agent -- can change when influenced by a $q$-panel, which consists of $q < N$ other agents chosen at random. If the $q$-panel is unanimous, meaning every agent in the $q$-panel is in the same state, the target agent takes their state, a scenario known as conformity.

In the absence of unanimity in the $q$-panel, the target agent will change its state with a constant probability, independent of the proportion of agents in adopted or unadopted states within the $q$-panel. If the target agent is currently  unadopted, it will switch to an adopted state with probability $\varepsilon_\uparrow$, and with the complementary probability $1 - \varepsilon_\uparrow$, it will remain unadopted. Conversely, if the target agent is already adopted, it will switch to the  unadopted state with probability $\varepsilon_\downarrow$, and with complementary probability $1 - \varepsilon_\downarrow$, it will stay adopted. Our model therefore includes two parameters $\epsilon_\uparrow$ and $\epsilon_\downarrow$, that accounts for the bias of an agent, depending on its current state. This is different from the idea of bias in partisan voter models, where an agent has inherent and fixed bias towards a specific opinion, irrespective of its current state \cite{masuda2010heterogeneous,llabres2023partisan}. 

\begin{figure}
\includegraphics[width=0.5\textwidth]{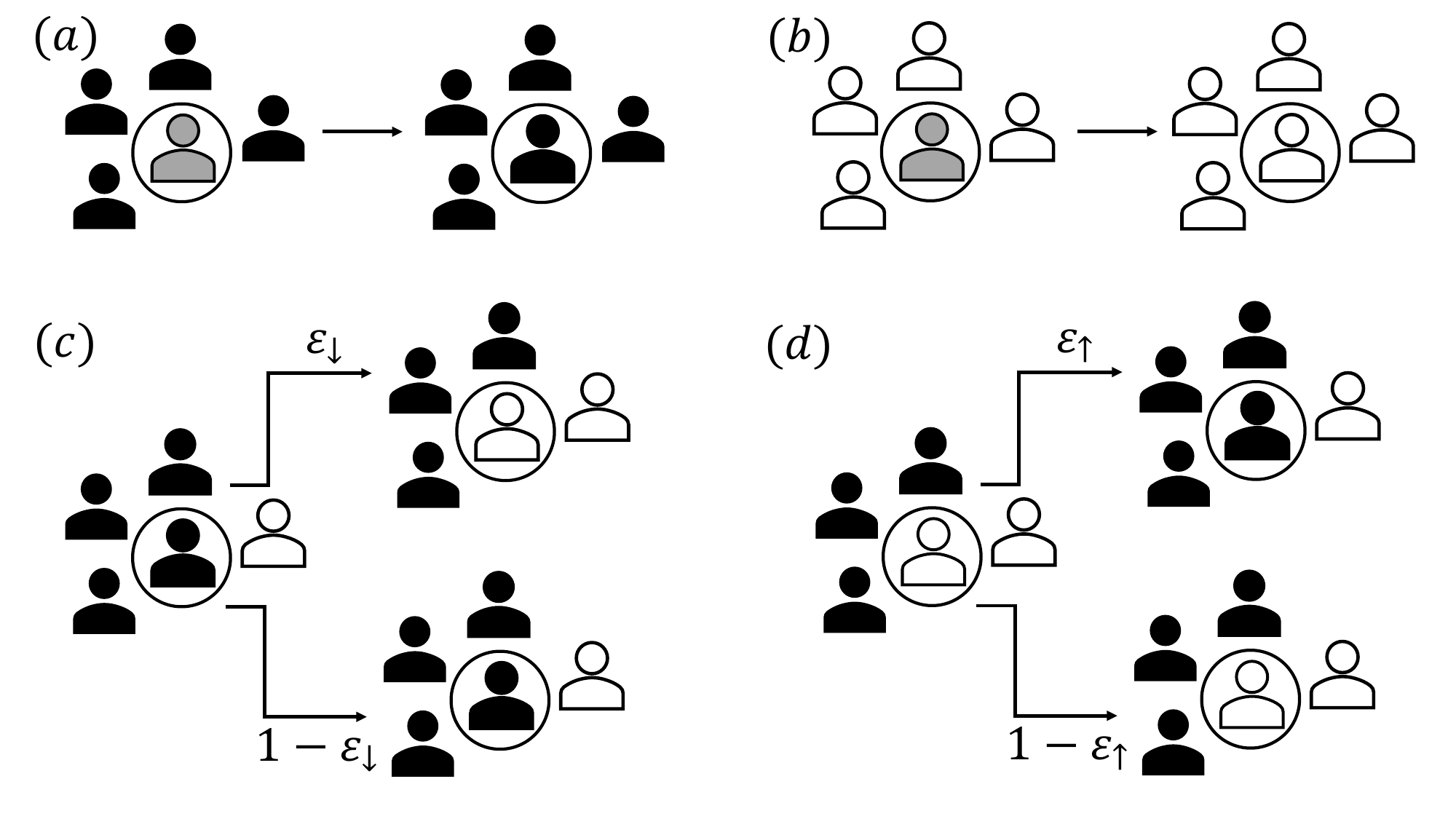}
\caption{Schematic diagram of our model illustrating possible scenarios in which a target agent (inside the circle) may change its state. Examples are provided for $q=4$. Black (white) agents represent adopted (unadopted) agents, while gray agents indicate agents in an arbitrary state. In scenarios (a) and (b), the agent's state changes irrespective of its previous opinion because the $q$-panel is unanimous. Cases (c) and (d) represent situations where the $q$-panel is not unanimous, meaning that the probability of changing state depends on the target agent's initial state: (c) the target agent is adopted and switches to an unadopted state with probability $\varepsilon_\downarrow$; otherwise, it retains its original state; (d) the target agent is unadopted and adopts with probability $\varepsilon_\uparrow$; otherwise, it retains its original state. These asymmetries introduce biases in decision-making that significantly impact the system's long-term behavior.}
\label{fig:diagram}
\end{figure}

The dynamics of our model for $q=4$ are shown in Fig. \ref{fig:diagram}. One time step in our model consists of $N$ elementary updates, where, in each elementary update, a target agent is randomly selected from the population of $N$ agents, and its state is updated. For $\epsilon_\uparrow = \epsilon_\downarrow$ it reduces to the original $q$-voter model, and for $\epsilon_\downarrow = 0$ to the model with mass media \cite{MUSLIM2024129358}.

\section{Results}\label{results}
\subsection{Aggregated description of the system}
\label{ana.res}
Let us denote by $N_{\uparrow}$ the number of adopted agents and by $N_{\downarrow}$ the number of unadopted agents. Since the total number of agents $N$ is fixed:
\begin{equation}
N_{\uparrow} + N_{\downarrow} = N.     
\end{equation}
Taking this into account, and considering the model on a complete graph, we can fully describe the state of the system by the aggregate variable $N_{\uparrow}$. As usual, it is more convenient to use intensive variables rather than extensive ones such as $N_{\uparrow}$. Therefore, following previous studies \cite{nyczka2012phase,jed:szn:19}, we introduce the concentration $c$ of adopted agents as:  
  \begin{equation}
    c=\frac{N_\uparrow}{N} \quad \rightarrow \quad 1-c=\frac{N_\downarrow}{N}.
\end{equation}
As usual, we can write the rate equation as:
\begin{equation}
    \frac{\Delta c}{\Delta t}=\gamma^+ - \gamma^-,
    \label{eq:dc_gamma_p_m_disc}
\end{equation}
where $\gamma^+$ is the probability of an event in which, during a time step $\Delta t = 1/N$, the number of adopted agents, $N_\uparrow$, increases by 1 (and simultaneously, $c$ increases by $\Delta c = 1/N$), and $\gamma^-$ is the probability of an event in which, during the same time step $\Delta t = 1/N$, the number of adopted agents decreases by 1 (and simultaneously, $c$ decreases by $\Delta c = 1/N$). For an infinite system size, $N \to \infty$, the time step $\Delta t \to 0$, and we can rewrite Eq. \eqref{eq:dc_gamma_p_m_disc} in a continuous form:
\begin{equation}
    \frac{d c}{d t}=\gamma^+ - \gamma^-.
    \label{eq:dc_gamma_p_m}
\end{equation}
Since the state of the system can change due to either the unanimity or non-unanimity of the $q$-panel, we can decompose $\gamma^+$ and $\gamma^-$ into the components corresponding to these two cases as
\begin{equation}
\gamma^{+}=\gamma^{+}_{u}+\gamma^{+}_{nu}, \quad \gamma^{-}=\gamma^{-}_{u}+\gamma^{-}_{nu},
\label{eq:gamma_pm}
\end{equation}
where the subscript $u$ refers to unanimity, while $nu$ denotes non-unanimity. With these notations, $\gamma^{+}_{u}$ represents the probability of an event in which, during an elementary update of duration
$\Delta t = 1/N$, the number of adopted agents, $N_\uparrow$, increases by 1 (and simultaneously, $c$ increases by $\Delta c = 1/N$ ) due to unanimity. Similarly, $\gamma^{+}_{nu}$ represents the probability of an event in which $N_\uparrow$ increases by 1 (and $c$ increases by $\Delta c$) due to non-unanimity. Likewise, $\gamma^{-}_{u}$ and $\gamma^{-}_{nu}$ represent the probabilities of events in which $N_\uparrow$ decreases by 1 (and $c$ decreases by $\Delta c$) as a result of unanimity and non-unanimity, respectively.  

The probabilities of events in which we randomly select a non-adopted/adopted agent and that it changes to the opposite state after interacting with a unanimous $q$-panel can be expressed as:
\begin{equation}
    \gamma^+_{u}= \frac{N_{\downarrow}}{N}\prod^{q-1}_{i=0} \frac{N_\uparrow-i}{N-1-i},
    \label{eq:gamma_plus_u}
\end{equation}
\begin{equation}
    \gamma^-_{u}= \frac{N_{\uparrow}}{N}\prod^{q-1}_{i=0} \frac{N_{\downarrow}-i}{N-1-i},
    \label{eq:gamma_minus_u}
\end{equation} 
where the terms before the product symbol represents the probabilities of events in which a target agent is selected that is unadopted or adopted, respectively, from a population of $N$ agents. Once this target agent is selected,  the probability of an event in which we choose an agent in the $q$-panel who is in the opposite state to the target, from the remaining population of $N - 1$ agents, is $N_\uparrow/(N - 1)$ or $N_\downarrow/(N-1)$ depending on the target's state. Since $q$ agents in the same state are needed, this selection process is repeated $q$ times, as represented by the product term.

The probabilities of events in which we randomly select a non-adopted/adopted agent and that it changes to the opposite state after interacting with a non-unanimous $q$-panel can be expressed as:
\begin{equation}
    \gamma^+_{nu}= \varepsilon_\uparrow\frac{N_{\downarrow}}{N}\left(1-\prod^{q-1}_{i=0} \frac{N_\uparrow-i}{N-1-i}-\prod^{q-1}_{i=0} \frac{N_{\downarrow}-1-i}{N-1-i}\right),
    \label{eq:gamma_plus_nu}
\end{equation}
\begin{equation}
    \gamma^-_{nu}= \varepsilon_{\downarrow}\frac{N_{\uparrow}}{N}\left(1-\prod^{q-1}_{i=0} \frac{N_\uparrow-1-i}{N-1-i}-\prod^{q-1}_{i=0} \frac{N_{\downarrow}-i}{N-1-i}\right).
    \label{eq:gamma_minus_nu}
\end{equation}
The terms within the parentheses denote the probability of an event in which a $q$-panel is selected that contains neither $q$ positive agents nor $q$ negative agents.

Now we substitute explicit expressions for $\gamma^+_{u}, \gamma^-_{u},\gamma^+_{nu}$ and $\gamma^-_{nu}$ given by Eqs.\eqref{eq:gamma_plus_u}-\eqref{eq:gamma_minus_nu} into Eq. \eqref{eq:gamma_pm}, and take the $N\rightarrow \infty $ limit to simplify the forms of $\gamma^+$ and $\gamma^-$, defined by Eq. \eqref{eq:gamma_pm}. As a result we get: 
\begin{eqnarray}
    \gamma^+ & = & (1-c)c^q +\varepsilon_\uparrow(1-c)(1-c^q-(1-c)^q), \nonumber\\
    \gamma^- & = & c(1-c)^q +\varepsilon_{\downarrow}c(1-c^q-(1-c)^q).
    \label{eq:gammas}
\end{eqnarray}

\begin{figure}
    \includegraphics[width=0.5\textwidth]{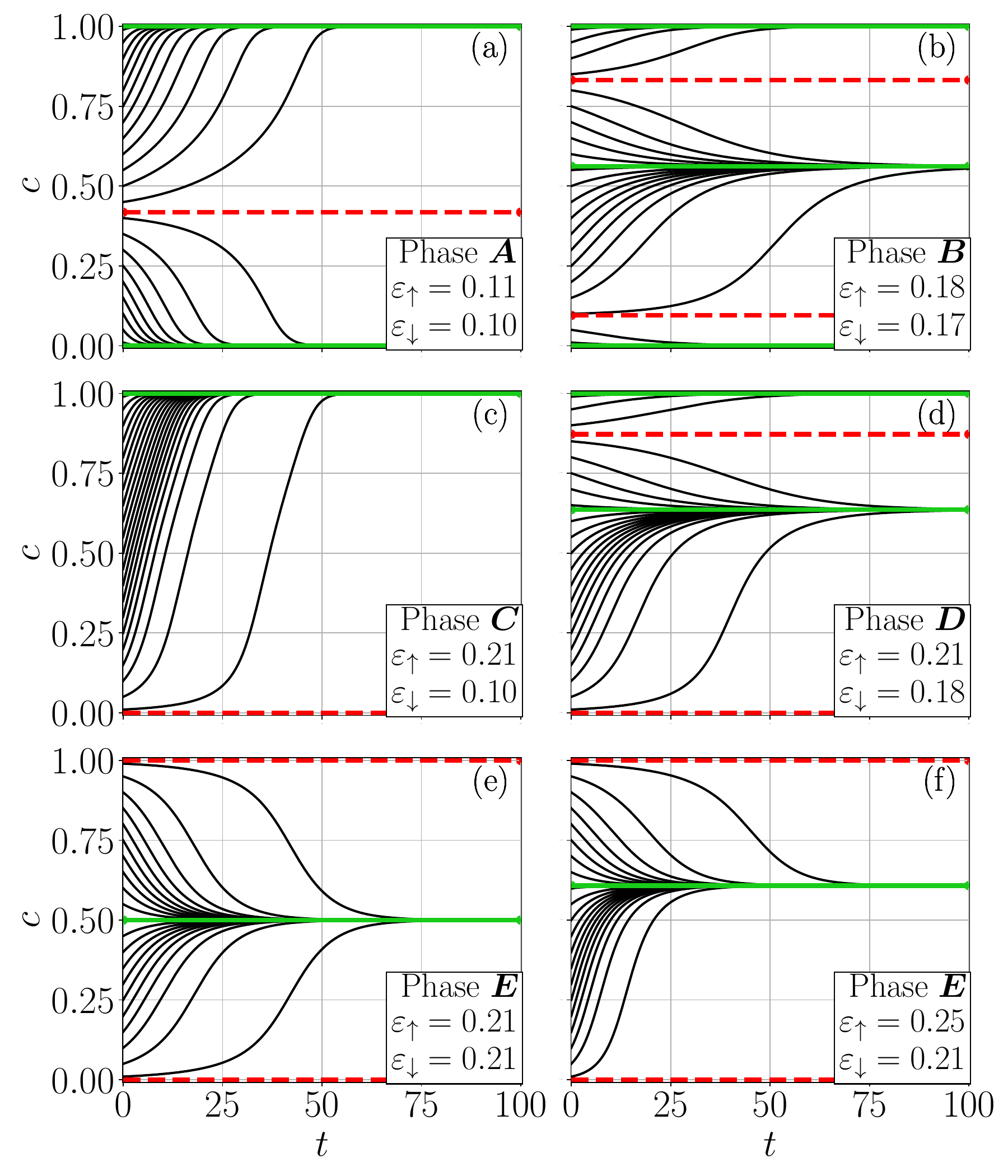}
    \caption{The temporal behavior of the concentration $c$ of adopted agents for $q=5$ and several values of $\varepsilon_\uparrow$ and $\varepsilon_\downarrow$ from different initial conditions. Trajectories (black solid line) are the numerical solutions of Eq. \eqref{eq:dc_dt}. The solid green lines indicate stable, whereas dashed red lines unstable fixed points of Eq. \eqref{eq:dc_dt} obtained from Eq. \eqref{eq:stationary}. The different panels correspond to different phases of the system, as indicated in the legends, where: (a) adopted or unadopted states are stable; (b) a partially adopted state emerges as an additional stable state; (c) only the adopted state is stable; (d) both adopted and a partially adopted state are stable; (e, f) only the partially adopted state is stable. These results illustrate how asymmetric adoption probabilities can lead to complex dynamical behavior.}
    \label{fig:traj}
\end{figure}

Inserting the probabilities given by Eq. \eqref{eq:gammas} to Eq. \eqref{eq:dc_gamma_p_m} we obtain:
\begin{eqnarray}
    \frac{dc}{dt} &=&(1-c)c^q - c(1-c)^q\nonumber \\  &+&(\varepsilon_\uparrow+\varepsilon_{\downarrow})\left (\frac{\varepsilon_\uparrow}{\varepsilon_\uparrow+\varepsilon_{\uparrow}} - c\right) \left(1-c^q-(1-c)^q\right). 
    \label{eq:dc_dt}
\end{eqnarray}
The above equation allows us to numerically study the time evolution of the system for arbitrary values of $\varepsilon_\uparrow$ and $\varepsilon_\downarrow$, as shown in Fig. \ref{fig:traj}. Note that the trajectories in Fig. \ref{fig:traj} were plotted for $\varepsilon_\uparrow \geq \varepsilon_\downarrow$, which corresponds to the situation where the adopted state is preferred. However, for $\varepsilon_\uparrow \leq \varepsilon_\downarrow$, the trajectories would look exactly the same if we examined the concentration $1-c$ of the unadopted agents instead of the adopted ones. Note also the solid green lines and the dashed red lines, which are fixed points (stationary states) of Eq. \eqref{eq:dc_dt} obtained from:
\begin{equation}
\frac{dc}{dt}=0. 
\label{eq:stationary}
\end{equation}
From Eq. \eqref{eq:dc_dt} the conditions for the fixed points are the following:
\begin{eqnarray}
\varepsilon_\uparrow &=&\frac{(1-c)c^q - c(1-c)^q}{(c-1)\left (1-c^q-(1-c)^q \right )} +\varepsilon_{\downarrow}\frac{  c}{1-c},\nonumber\\
\varepsilon_{\downarrow} &=&\frac{(1-c)c^q - c(1-c)^q}{c \cdot \left (1-c^q-(1-c)^q \right )}+\varepsilon_{\uparrow}\frac{1-c}{c}.
\label{eq:stationary_epsilon}
\end{eqnarray}

From Eqs. \eqref{eq:stationary_epsilon}, we can obtain several results that have been reported earlier in the literature:  for $\varepsilon_\uparrow=\varepsilon_{\downarrow}$ the model boils down to the  original $q$-voter by Castellano et. al. \cite{qvm_castellano}:
\begin{equation}
    \varepsilon_{\uparrow}=
    \frac{c(1-c)^q - (1-c)c^q}{(1-2c)(1-c^q-(1-c)^q)},
\end{equation}
and  for $\varepsilon_\downarrow = 0$ the model with mass media \cite{MUSLIM2024129358}
\begin{equation}
     \varepsilon_{\uparrow}
    =\frac{c(1-c)^q - (1-c)c^q}{(1-c)(1-c^q-(1-c)^q)}.
\end{equation} On the other hand, our model with $\varepsilon_\uparrow+\varepsilon_\downarrow=1$ is identical to that studied by Mullick and Sen \cite{mullick2025social} only when $q=2$.

%It is worthwhile to note whether the voter model type behavior will occur for any parameter range. 
Following \cite{qvm_castellano}, one can obtain the equation obeyed by $P(c)$, the probability that the fraction of adopted agents is $c$ as
\begin{align}
\frac{\partial P(c,t|c^\prime, t^\prime)}{\partial t^\prime} &= (\gamma^+ - \gamma^-) \frac{\partial P(c,t|c^\prime, t^\prime)}{\partial c^\prime} \nonumber \\
& + \frac{(\gamma^+ + \gamma^-) }{2N} \frac{\partial^2 P(c,t|c^\prime, t^\prime)}{\partial c^{\prime 2}},
\label{probeq}
\end{align}
where the expressions for $\gamma^+$ and $\gamma^-$ are given by Eq. \eqref{eq:gammas}.
The terms on the right correspond to a drift and diffusion respectively and are to be considered as functions of the primed variables.
For the voter model, the drift term vanishes. For $q=1$, for any $\varepsilon_\uparrow$ and $\varepsilon_\downarrow$, one gets voter model trivially. Checking for $q=2$, we obtain the 
condition for voter model behavior to prevail for 
\begin{equation}
c(1-c)\left(c-\frac{1}{2} +\varepsilon_\uparrow(1-c) -   \varepsilon_\downarrow c\right) = 0,
\end{equation}
which can be only satisfied for $\varepsilon_\uparrow= \varepsilon_\downarrow= 1/2$.
For larger values of $q$, the analysis becomes less simple to handle unless $\varepsilon_\uparrow = \varepsilon_\downarrow$ (for which the linear voter model ceases to exist for $q > 2$). However, this approach is useful to study the exit probability as well, shown later in this paper.

\subsection{Linear stability analysis}
A better understanding of the model involves not only finding the fixed points, but also analyzing their stability. Finding the analytical form for all solutions of Eq. \eqref{eq:dc_dt} for any $\varepsilon_\uparrow$, $\varepsilon_\downarrow$, and $q$ is very difficult, if not impossible. Therefore, in the next section, we will study the system numerically. However, several solutions are straightforward.

Note that the target agent needs at least one adopted agent  in its neighborhood to adopt. Therefore, a state where all agents are unadopted ($c=0$) is a steady state. A similar argument could be used to explain that a state where all agents are adopted ($c=1$) is also a steady state. It is also easy to see  that $c=0$ and $c=1$ are indeed solutions of Eq. \eqref{eq:dc_dt} for any value of $\varepsilon_\uparrow$, $\varepsilon_\downarrow$, and $q$, because $(1-c)c^q - c(1-c)^q = 0$ and $1-c^q-(1-c)^q = 0$ for both $c=0$ and $c=1$.

To check when these solutions would be stable, we need to determine the criteria for which:
\begin{equation}
    \frac{d}{dc}\left(\frac{dc}{dt}\right)<0.
\end{equation}
First we calculate the derivative from Eq. \eqref{eq:dc_dt} as
\begin{eqnarray}
 %\begin{aligned}
\frac{d}{dc}\left(\frac{dc}{dt}\right) &=& \ q \left( (1-c)c^{q-1} + c(1-c)^{q-1} \right) - c^q - (1-c)^q\nonumber \\
&+&  (\varepsilon_\uparrow + \varepsilon_\downarrow) \Bigg[ q \left( \frac{\varepsilon_\uparrow}{\varepsilon_\uparrow + \varepsilon_\downarrow} - c \right)\nonumber \\
&\times &  \left( (1-c)^{q-1} - c^{q-1} \right) \nonumber\\ &-&  (1 - c^q - (1-c)^q) \Bigg].
\label{eq:dc_dc_dt}
%\end{aligned}
\end{eqnarray}

From the above equation, we see that the fixed point $c=0$ is stable for
\begin{equation}
      \varepsilon_\uparrow   < \frac{1}{q},
\end{equation}
whereas the fixed point $c=1$ is stable for
\begin{equation}
\varepsilon_\downarrow<\frac{1}{q}.
\end{equation}
The above conditions hold for our generalized model. However, if we reduce the model to the original $q$-voter model by assuming that $\varepsilon_\uparrow = \varepsilon_\downarrow$, we obtain another stationary state for $c=1/2$. The criteria for this state to be stable could be obtained from Eq. \eqref{eq:dc_dc_dt} as
\begin{equation}
  \varepsilon_\uparrow  >
  \frac{q-1}{2^q-2},
\end{equation} which agrees with the result shown in \cite{qvm_castellano}.

\subsection{Phase diagram} \label{time_traj}
\begin{figure*}
    \includegraphics[width=\textwidth]{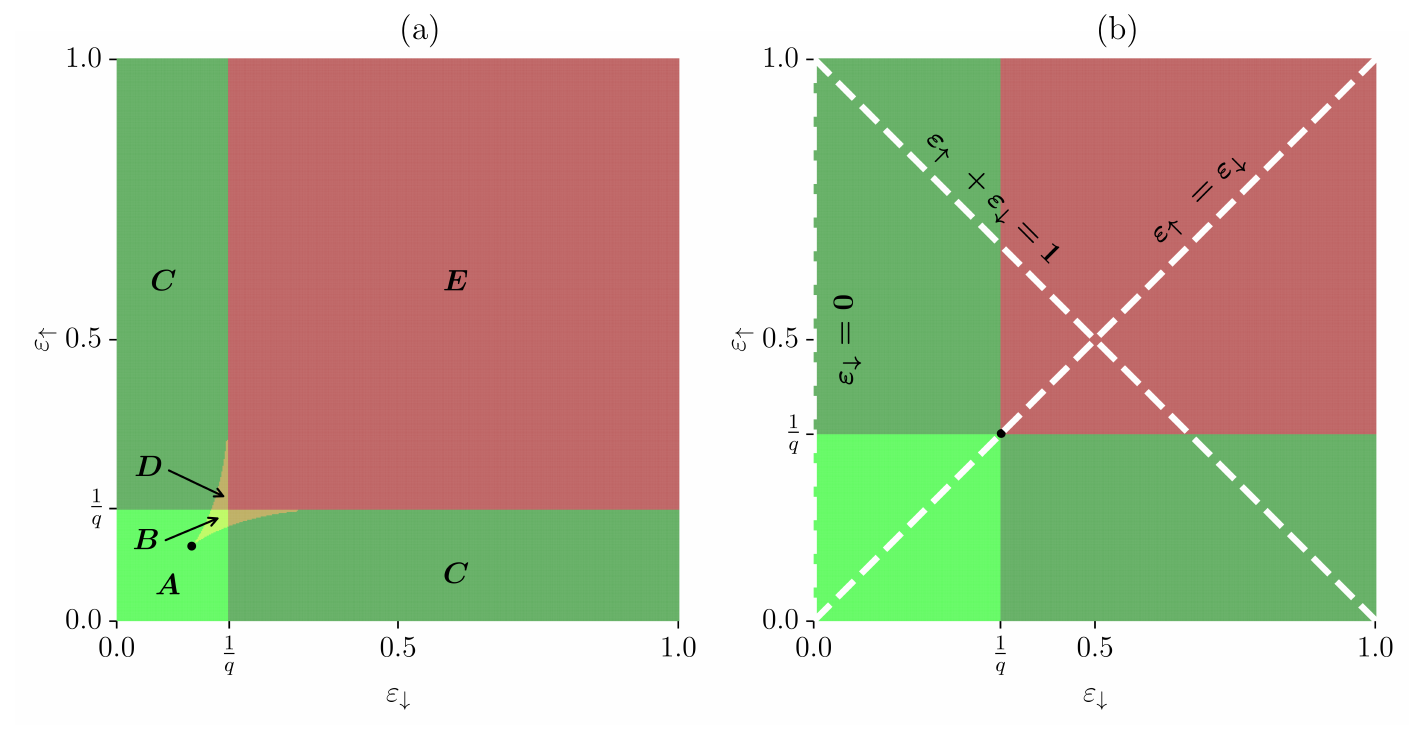}
    \caption{Phase diagrams for (a) $q=5$ and (b) $q=3$. The temporal behavior of the system in each of the phases $\boldsymbol{A}$ -- $\boldsymbol{E}$ is shown in Fig. \ref{fig:traj}. The values $\varepsilon_\downarrow=1/q$ and $\varepsilon_\uparrow=1/q$, which define the boundaries between the phases, separate the stability from the unstability of the stationary states $c=1$ and $c=0$. The black dot marks the point $\varepsilon_\uparrow = \varepsilon_\downarrow = (q-1)/(2^{q}-2)$, above which the state $c=1/2$ becomes stable. In the left panel the capital letters denote phases, while in the right panel the white dashed lines denote special cases of our generalized model: $\varepsilon_\downarrow=\varepsilon_\downarrow$ corresponds to the original $q$-voter model \cite{qvm_castellano} and $\varepsilon_\downarrow=0$ to the new model with mass media \cite{MUSLIM2024129358}. Comparing panels (a) and (b), we observe two things. First, phase $\mathbf{D}$ was not present in previous models. Second, phases $\mathbf{B}$ and $\mathbf{D}$ are absent for $q = 3$ but present for $q = 5$. In fact, they appear only for an influence group size of $q > 3$. This means that the coexistence of partially adopted and adopted states, which occur only in phases $\mathbf{B}$ and $\mathbf{D}$, is possible only for larger influence groups, which is particularly interesting from a social perspective.}
    \label{fig:map}
\end{figure*}

\begin{figure}
    \includegraphics[width=0.5\textwidth]{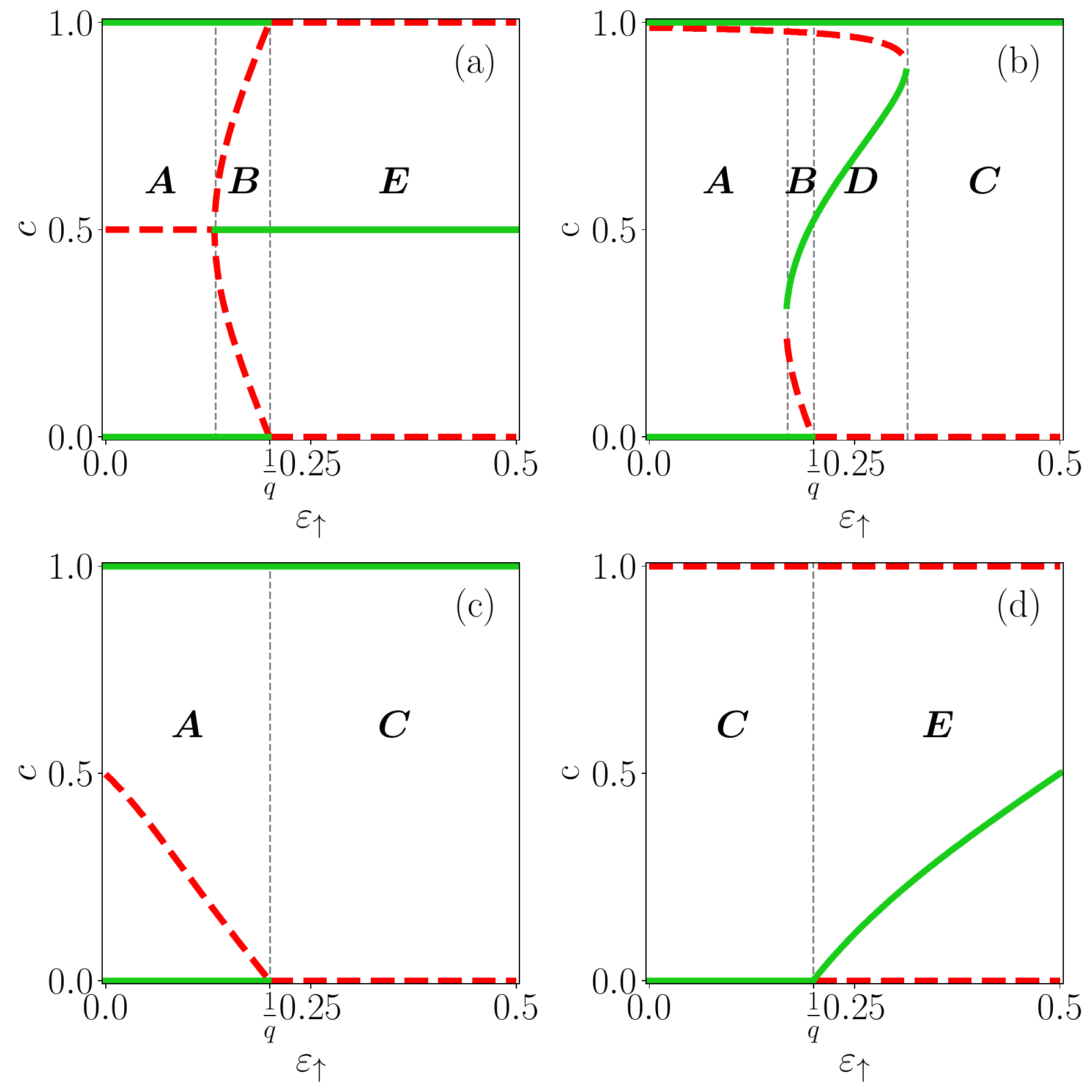}
    \caption{Concentration of adopted agents in stationary states for $q=5$ as a function of $\varepsilon_\uparrow$ for (a) $\varepsilon_\downarrow=\varepsilon_\uparrow$, which corresponds to the original $q$-voter model \cite{qvm_castellano}, (b) $\varepsilon_\downarrow=0.195<1/q$, which allows to pass through 4 different phases (the novel behavior of our generalized model), for comparison see Fig. \ref{fig:map} (a), (c) $\varepsilon_\downarrow = 0$, which corresponds the model with mass media \cite{MUSLIM2024129358} and (d) $\varepsilon_\uparrow+\varepsilon_\downarrow=1$, with complementary biases for adopted and unadopted states. The solid green lines indicate stable states. The dashed red lines indicate unstable states. The dashed gray vertical line marks the transition between phases, and the capital letters indicate the phases that occur in the system. This diagram, although showing only stationary states, allows us to observe the dependence on initial conditions. If the system starts with an initial concentration of adopters below a certain dashed line, it will evolve toward the final concentration on the nearest solid line below the initial point. Conversely, if the system starts with an initial concentration above a certain dashed line, it will move toward the nearest solid line above it, reaching a different final concentration.}
    \label{fig:stac}
\end{figure}

As seen in Fig. \ref{fig:traj} there are three types of the fixed points:
\begin{itemize}
    \item $c=1$: adopted state
    \item $c=0$: unadopted state
    \item  $0 < c < 1$: partially adopted state.
\end{itemize}
Each of the above states can be a stable solution of Eq. \eqref{eq:stationary}, indicated by the solid green line in Fig. \ref{fig:traj}, or unstable, indicated by the dashed red line in Fig. \ref{fig:traj}. Based on the information of how many solutions of Eq. \eqref{eq:stationary} exist, and
which of these states are stable and which are not, we determine the phase of the system and construct the phase diagrams as shown in Fig. \ref{fig:map}. For our model, we can distinguish five phases, labeled $\boldsymbol{A}$, $\boldsymbol{B}$, $\boldsymbol{C}$, $\boldsymbol{D}$, and $\boldsymbol{E}$, see also  Figs. \ref{fig:traj} -- \ref{fig:stac}: 
\begin{itemize}
    \item[$\boldsymbol{A}$]: there are two stable states, unadopted ($c=0$) and adopted ($c=1$), and an unstable partially adopted state. For $\varepsilon_\uparrow > \varepsilon_\downarrow$, this unstable state lies below $c=1/2$, meaning that the adopted state is preferred. Specifically, the range of initial conditions leading to the adopted state is larger than that leading to the unadopted state. As $\varepsilon_\uparrow$ increases, this unstable state moves toward $c=0$. When $\varepsilon_\uparrow$ exceeds $1/q$, the stable state $c=0$ disappears, and the system enters phase $\boldsymbol{C}$, as shown in Fig. \ref{fig:traj}(c). 
    \item[$\boldsymbol{B}$]: there are three stable states (unadopted, adopted, and partially adopted), as well as two partially adopted unstable states. 
    \item[$\boldsymbol{C}$]: there is only one stable and one unstable state. For $\varepsilon_\uparrow > 1/q > \varepsilon_\downarrow$, the adopted state is stable and unadopted unstable, and $\varepsilon_\downarrow > 1/q >\varepsilon_\uparrow$ vice versa, as shown in Fig. \ref{fig:map}. 
    \item[$\boldsymbol{D}$]: there are two stable states  and two unstable states. For $\varepsilon_\uparrow > 1/q > \varepsilon_\downarrow$, adopted and partially adopted states are stable, while unadopted and another partially adopted states are  unstable. Obviously, for $\varepsilon_\downarrow > 1/q >\varepsilon_\uparrow$ we have a complementary situation, as shown in Fig. \ref{fig:map}. 
    \item[$\boldsymbol{E}$]: there is only one stable solution, which is actually a partially adopted state. Both, adopted and unadopted states are unstable. In Figs. \ref{fig:traj}(e) and (f) we have presented two different variants of phase $\boldsymbol{E}$ with $\varepsilon_{\uparrow}=\varepsilon_{\uparrow}>1/q$ (symmetric case) and $\varepsilon_{\uparrow}>\varepsilon_{\downarrow}>1/q$ (asymmetric case) respectively.
\end{itemize}

We summarize the above results in Fig. \ref{fig:map}, which presents the phase diagrams in the $(\varepsilon_\uparrow ,\varepsilon_\downarrow)$ plane. The left panel of Fig. \ref{fig:map} shows the phase diagram for $q=5$, corresponding to Figs. \ref{fig:traj} and \ref{fig:stac}, with labeled phases. In the right panel  the phase diagram for $q=3$ is shown, illustrating how the size of the influence group affects the system's behavior. Additionally, in this panel, we include two dashed white lines representing the parameter values of $\varepsilon_\uparrow$ and $\varepsilon_\downarrow$ from earlier models \cite{qvm_castellano,MUSLIM2024129358}. We have chosen not to place these lines in the left panel because it would make the diagram unreadable. Comparing two panels of Fig. \ref{fig:map}, we see that the $\boldsymbol{D}$ phase was not present in the previous models \cite{qvm_castellano,MUSLIM2024129358}.

The phase diagrams in Fig. \ref{fig:map} are clearly symmetric with respect
to the line $ \varepsilon_\uparrow=\varepsilon_\downarrow $, which represents the original $q$-voter model \cite{qvm_castellano}. This symmetry arises because $ \varepsilon_\uparrow $ and $ \varepsilon_\downarrow $ contribute equally to the dynamics. 

Let us first focus on Fig. \ref{fig:map}(a), which shows the phase diagram for $q=5$. When $\varepsilon_\uparrow= \varepsilon_\downarrow=0$, the system is in phase $\boldsymbol{A}$. If we increase $\varepsilon_\uparrow$ to $ (q-1)/(2^{q}-2) $ while maintaining $\varepsilon_\uparrow=\varepsilon_\downarrow$, we reach phase $\boldsymbol{B}$, where a partially adopted steady state becomes possible. Beyond the lines $\varepsilon_\uparrow=1/q$ and $\varepsilon_\downarrow=1/q$, the system loses one of the stable fully adopted states. Specifically, phase $\boldsymbol{A}$, upon losing one fully adopted state, transitions to phase $\boldsymbol{C}$, while phase $\boldsymbol{B}$, upon losing one fully adopted state, transitions to phase $\boldsymbol{D}$. Each of these transitions can occur in two distinct ways, as shown in Fig. \ref{fig:traj}(a), depending on whether $\varepsilon_\uparrow>1/q>\varepsilon_\downarrow$ or $\varepsilon_\downarrow>1/q>\varepsilon_\uparrow $. However, when $\varepsilon_\uparrow>1/q$ and $\varepsilon_\downarrow>1/q $, the system reaches phase $ \boldsymbol{E}$, where neither a fully adopted nor a fully unadopted state is possible as a steady state, and the only stable solution is a partially adopted state. These transitions between the phases are easier to understand if we plot the stationary valued of adopted $c$ as a function of only one model's parameter, for example $\varepsilon_\uparrow$, as shown in Fig. \ref{fig:stac}. 

The phase diagram shown in Fig. \ref{fig:map}(a) follows the same pattern for all values of $q>3$, with the width of phase $\boldsymbol{C}$ decreasing as $q$ increases. However, for $q=2$ and $q=3$, phases $\boldsymbol{B}$ and $\boldsymbol{D}$ are absent, as shown in Fig. \ref{fig:map}(b). This is because the inequality $ (q-1)/(2^{q}-2)<1/q$ does not hold for $q=2$ and $q=3$, meaning that the coexistence of adopted/unadopted and partially adopted steady states is not possible.

\subsection{Exit probability}
In models of opinion dynamics, an important quantity of interest is the exit probability $E(c_0)$, which is defined as the probability that the system will reach an adopted state ($c=1$) from an initial fraction of adopted agents equal to $c_0$. In our model, an adopted state is stable only for some values of the model parameters, while it is unstable for others. In the latter case the adopted state will never be reached if the system size $N$ is large. However, if the system size is small, the unstable state can be reached due to finite size fluctuations. Once the adopted state is entered, it will never be left, since $c=1$ is an absorbing state by the definition of the model. This motivated us to determine the exit probability $E(c_0)$ for small systems within Monte Carlo simulations and additionally calculate it analytically. 

\begin{figure*}
    \includegraphics[width=\textwidth]{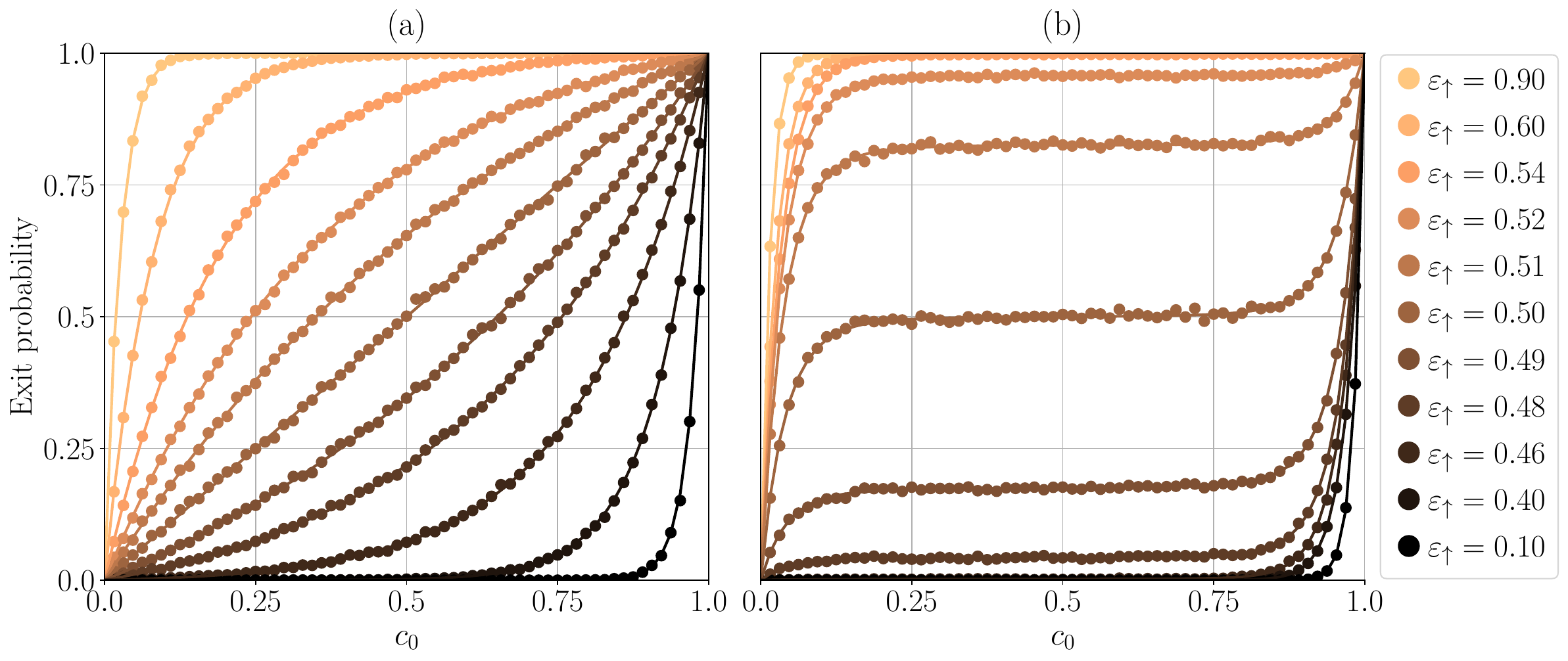}
    \caption{Exit probability $E(c_0)$ as a function of initial concentration of adopted  agents  $c_0$ for several values of $\varepsilon_\uparrow$, where $\varepsilon_\uparrow + \varepsilon_\downarrow =1$, for (a) $q=2$ and (b) $q=3$. The dots represent outcome of the Monte Carlo simulations for a complete graph of size $N = 64$, averaged over $10^4$ realizations. Lines represent analytical results: for $q=2$ they can be obtained from the closed-form expression \eqref{exitq2} or from the numerical solution of Eq. \eqref{eq:exit_markov}, while for $q \ge 3$ only the numerical solution of Eq. \eqref{eq:exit_markov} are possible. For $q=2$, the exit probability follows an expected linear trend for $\varepsilon_\uparrow=0.5$, but deviates as asymmetry increases. For $q=3$, a plateau appears in the exit probability, indicating that increasing initial adoption does not necessarily increase the likelihood of full adoption in small systems. This effect suggests that group influence size plays an important role in determining collective decision outcomes.}
    \label{fig:exit_prob}
\end{figure*}

We focus on the case $\varepsilon_\uparrow+\varepsilon_\downarrow=1$ and start with Monte Carlo simulations to determine the exit probability for several values of $\varepsilon_\uparrow$ and $q$ for the system of size $N=64$; the results are shown in Fig. \ref{fig:exit_prob}. 

In Fig. \ref{fig:exit_prob}(a), which presents the results for $q=2$, we observe that for $\varepsilon_\uparrow=1/2$, the exit probability follows the line $E(c_0) = c_0$. Additionally, we see that the probability of reaching the adopted state increases as $\varepsilon_\uparrow$ increases for any $c_0$, though the manner in which it increases depends strongly on $\varepsilon_\uparrow$. For instance, for small values of $\varepsilon_\uparrow$ (e.g., $\varepsilon_\uparrow=0.1$), only high values of $c_0$ ensure that the adopted state is reached, while most initial conditions lead to the unadopted state, which is an expected result. However, what is unexpected is the influence of $q$ on the exit probability, as evidenced by the significant difference between the exit probabilities for $q=2$ and $q=3$.  

In Fig. \ref{fig:exit_prob}(b), which presents the results for $q=3$, we observe a plateau region resulting from the existence of a stable partially adopted state, covering the range from $\varepsilon_\uparrow=0.4$ to $\varepsilon_\uparrow=0.6$. The emergence of this plateau can be understood from Fig. \ref{fig:map}(b), where we highlight the line $\varepsilon_\uparrow+\varepsilon_{\downarrow}=1$. Once $\varepsilon_\uparrow$ exceeds $1/q$, the system enters phase $\boldsymbol{E}$, where the only stable state is a partially unadopted state. This situation does not occur for $q=2$, as the $\varepsilon_\uparrow+\varepsilon_{\downarrow}=1$ line never intersects phase $\boldsymbol{E}$. For $q>3$, the exit probability results remain qualitatively similar to those for $q=3$.  

The exit probability can be obtained not only from simulation but also analytically. From Eq. \eqref{probeq}, at the steady state:
\begin{equation}
\frac{\partial^2 E(c_0)} {\partial c_0^2}
        + {2N}\bigg(\frac{\gamma^+ - \gamma_-}{\gamma^+ + \gamma_-}\bigg)\frac{\partial E(c_0)}{\partial c_0}  = 0.
% \label{exiteq}
\end{equation}
The solution, satisfying $E(0) = 0$  and $E(1) = 1$ can be written as
\begin{equation}
        E(c_0) = \frac{\int_0^{c_0} \exp (-\int g(c_0)dc_0) dc_0}{\int_0^1 \exp (-\int g(c_0)dc_0) dc_0},
\end{equation}
where
\begin{equation}
g(c_0) =  2N \bigg(\frac{\gamma^+ - \gamma_-}{\gamma^+ + \gamma_-}\bigg).
\end{equation}
The $q=2$ case can be handled easily. For the voter model behavior for $q=2$, as noted earlier, $g(c_0) = 0$ leads to  $E(c_0) = c_0$ which is obtained for $\varepsilon_\uparrow = \varepsilon_\downarrow = 1/2$. In general,  
\begin{equation}
g(c_0) = \frac{c_0(1-(\varepsilon_\uparrow + \varepsilon_\downarrow)) +\varepsilon_\uparrow -\frac{1}{2}}{c_0(\varepsilon_\downarrow - \varepsilon_\uparrow) +\varepsilon_\uparrow +\frac{1}{2}}.
\end{equation}
For $\varepsilon_\uparrow + \varepsilon_\downarrow =1$, we have further simplification and the solution is (for $c_0 \neq 1/2$)
\begin{equation}
        E(c_0) = \frac{\left(c_0(1-2\varepsilon_\uparrow)+\left(\varepsilon_\uparrow+\frac{1}{2}\right)\right)^{N+1} -\left(\varepsilon_\uparrow+\frac{1}{2}\right)^{N+1}}
        {\left(\frac{3}{2}-\varepsilon_\uparrow\right)^{N+1} - \left(\varepsilon_\uparrow+\frac{1}{2}\right)^{N+1}}.
\label{exitq2}
\end{equation}
 Eq. \eqref{exitq2} provides an analytical solution for the exit probability in the case of $q=2$. Cases for higher values $q>2$ can be obtained numerically using the Markov Chain approach, presented below.

Since our system is fully homogeneous, the possible number of states of the system is $N+1$ ($N_\uparrow = 0, 1, \dots, N$, or equivalently $c = 0, 1/N, \dots, 1$). This allows us to represent our system as an $(N+1)$-state Markov chain\cite{Ban:Sve:15}. The transition probabilities between states can be derived from Eq. \eqref{eq:gammas}. We treat these probabilities as a function of $c$, where the concentration in state $k$ is $c = k/N$.  

The probabilities of events are as follows: for the transition from state $k$ to $k+1$, it is denoted by $\gamma^+(k)$; for the transition from state $k$ to $k-1$, it is denoted by $\gamma^-(k)$; and remaining in state $k$ is denoted by  $\gamma^0(k) = 1 - \gamma^+(k) - \gamma^-(k)$. Thus, the transition matrix for our Markov chain is the following:
\begin{equation*}
\renewcommand{\arraystretch}{1.5}
\mathit{P} =
\begin{bmatrix}
1 & 0 & 0 &0 & \cdots & 0 \\
\gamma^-(\frac{1}{n}) & \gamma^0(\frac{1}{n}) &\gamma^+(\frac{1}{n}) &0 & \cdots & 0 \\
0 & \gamma^-(\frac{2}{n}) & \gamma^0(\frac{2}{n}) &\gamma^+(\frac{2}{n}) &\cdots & 0 \\
\vdots & \vdots & \ddots& \ddots & \ddots & \vdots \\
0 & \cdots  &\gamma^-(\frac{n-2}{n}) & \gamma^0(\frac{n-2}{n}) &\gamma^+(\frac{n-2}{n}) & 0\\
0 & \cdots & 0& \gamma^-(\frac{n-1}{n}) & \gamma^0(\frac{n-1}{n}) &\gamma^+(\frac{n-1}{n})   \\
0 & \cdots & 0& 0 & 0 & 1
\end{bmatrix}
\end{equation*}
For the exit probabilities, we need the fundamental matrix. However, we cannot construct it if there are absorbing states in the system.  
Therefore, we construct the transition matrix for a new Markov chain, excluding the absorbing states $k=0$ and $k=N$:

\begin{equation*}
\renewcommand{\arraystretch}{1.5}
\mathit{Q} =
\begin{bmatrix}
\gamma^0(\frac{1}{n}) &\gamma^+(\frac{1}{n}) & 0 &0 & \cdots & 0 \\
\gamma^-(\frac{2}{n}) & \gamma^0(\frac{2}{n}) &\gamma^+(\frac{2}{n}) &0 & \cdots & 0 \\
0 & \gamma^-(\frac{3}{n}) & \gamma^0(\frac{3}{n}) &\gamma^+(\frac{3}{n}) &\cdots & 0 \\
\vdots & \vdots & \ddots& \ddots & \ddots & \vdots \\
0 & \cdots  &\gamma^-(\frac{n-3}{n}) & \gamma^0(\frac{n-3}{n}) &\gamma^+(\frac{n-3}{n}) & 0\\
0 & \cdots & 0& \gamma^-(\frac{n-2}{n}) & \gamma^0(\frac{n-2}{n}) &\gamma^+(\frac{n-2}{n})   \\
0 & \cdots & 0& 0 & \gamma^-(\frac{n-1}{n}) & \gamma^0(\frac{n-1}{n})
\end{bmatrix}
\end{equation*}
Now we can obtain the fundamental matrix by solving the following equation numerically:
\begin{equation}
    \mathit{F}=(\boldsymbol{1}- \mathit{Q})^{-1}.
\end{equation}
Here $\boldsymbol{1}$ is the identity matrix.  The elements of the fundamental matrix can be interpreted as the expected number of visits to a transient state.  Absorbing states do not appear in the matrix $\mathit{Q}$, but instead, we take state $k=N-1$ instead of the absorbing state $k=N$ and include information about the transition probability from state $k=N-1$ to state $k=N$, $\gamma^+(\frac{n-1}{n})$. Therefore, we construct the following matrix:
\begin{equation*}
\renewcommand{\arraystretch}{1.5}
\mathit{R} =
\begin{bmatrix}
0  \\
0\\
 \vdots  \\
0\\
 \gamma^+(\frac{n-1}{n})
\end{bmatrix}
\end{equation*}

By solving the following equation:  
\begin{equation}  
    \mathit{F} \cdot \mathit{R} = \left[ E\left(\frac{1}{N}\right), \dots, E\left(\frac{N-1}{N}\right) \right] \label{eq:exit_markov}
\end{equation}  
we obtain a vector containing the exit probabilities for all initial conditions that are not absorbing states. Due to the fact that the exit probability defines reaching state $c=1$, i.e., $k=N$, the exit probability for absorbing states is $E(0) = 0$ and $E(N/N) = 1$, regardless of the parameters $q, \varepsilon_\uparrow, \varepsilon_\downarrow$. The results from Eq. \eqref{eq:exit_markov} are shown in Fig. \ref{fig:exit_prob}, where they are compared with the results from the simulations. 
We can see that the equations match well with the results from simulations for $q = 2$ and $q = 3$.
In both cases, the quantitative divergence increases as the system becomes more asymmetric, i.e., the further we move from $\varepsilon_\uparrow = 1/2$.

\section{Discussions}\label{sec:dis}
In this paper we have introduced a generalized $q$-voter model. Social influence is modeled in the same way as in the original one \cite{qvm_castellano}: in the case of a unanimous $q$-panel, an agent adopts the state of the $q$-panel. The difference between the generalized and the original model occurs when the $q$-panel is not unanimous. In such a case, we assume that the probability $\varepsilon_\uparrow$ of going from an unadopted state to an adopted state is not equal to the probability $\varepsilon_\downarrow$ of going from an adopted state back to an unadopted state. For $\varepsilon_\downarrow=\varepsilon_\uparrow$ the model reduces to the original $q$-voter model \cite{qvm_castellano} and for $\varepsilon_\downarrow=0$ to the recently introduced mass media model \cite{MUSLIM2024129358}. 

The motivation for this generalization comes from two sources. First, it can be understood as a form of sequential bias, as an agent's current opinion influences its updated opinion. Second, it can be seen as a situation where adopted and unadopted states offer additional benefits or are somehow promoted, though they are not equivalent. Such situations frequently occur in real markets, for example, during the diffusion of innovation or in duopoly markets, where both products are actively promoted. 

The generalized asymmetric model exhibits rich behavior, but two results  are particularly interesting and have not been observed in previous works. First, a new phase $\boldsymbol{D}$ appears for $q > 3$, in which two stable and two unstable stable steady states exist: for $\varepsilon_\uparrow>\varepsilon_\downarrow$ adopted and partially adopted states are stable. It means that for any initial conditions $c_0$ the system reaches a partially adopted state, however to reach an adopted state, the initial condition of adopted must exceed a certain critical value. Similarly the oscillating external propaganda field in Gimenez et. al \cite{gimenez2023contrarian} prevents permanent consensus (adoption or non-adoption) and instead drives the system into a regime of sustained opinion oscillations or dynamic equilibria, showing how time-dependent biases can counteract the effects of asymmetry. However, this is in contrast with Galam et. al. \cite{galam2020asymmetric} where once asymmetry is introduced, consensus is effectively restored.

The second result is a very unique form of exit probability that appears for $q \ge 3$ in the case of small systems, and such systems are realistic when we consider social groups, for example organizations. In this case, we observe a broad plateau region, indicating that a larger fraction of initially adopted (or convinced) people does not lead to a higher probability of adoption of the entire system. In the language of opinion dynamics, we could also say that the larger number of people with positive opinions does not affect the probability of reaching a final positive consensus. On the other hand, for $q<3$ the exit probability has an expected form, e.g. for the symmetric case $\varepsilon_\uparrow = \varepsilon_\downarrow = 1/2$ it increases linearly with $c_0$. Notably, our model produces identical results with that of Mullick and Sen \cite{mullick2025social} for exit probability $E(x)$ when $q=2$. However the $E(x)$ results of their model remains qualitatively similar for any $q\geq 2$, which is clearly different than what we have found for $q>2$.

Therefore, to summarize our results show how even small asymmetries can lead to surprising collective behavior. The emergence of a stable partially adopted state for $q>3$ and the non-trivial exit probability plateau in small systems suggest that collective adaptation is not solely determined by initial support, but depends on multiple factors such as the size $q$ of the influence group and the size of the total social group making the collective decision. This is an interesting result from a social perspective, as the role of group size in collective action is one of the most important topics in social science \cite{Est:Deb:01}. There is long-standing empirical evidence that the size of a group is the best predictor of its level of collective action \cite{Oli:Mar:88}. However, whether the larger groups influence this collective action positively or negatively depends on the specific situation. The results of our generalized model show that both the size of the influence group $q$ and the size of the entire group $N$ that is involved in the collective adaptation actually matter.
Therefore, although the model is highly simplified, it provides useful insights into opinion dynamics and collective decision making. Future research could examine how these effects change under more realistic conditions, such as external influences, network structures, or strategic interventions, to better understand the mechanisms that drive collective adaptation, which is needed now more than ever \cite{Galesic2023}.

\begin{acknowledgments}
This research was partially funded by the National Science Centre, Poland Grant 2019/35/B/HS6/02530 (MD and KSW). PS thanks the financial support from Council of Scientific \& Industrial Research (CSIR), India (file no. 03/1495/23/EMR-II), as well as acknowledges the hospitality provided by Wrocław University of Science and Technology.
\end{acknowledgments}

\section*{Author Declarations}
The authors have no conflicts to disclose.

\section*{Data Availability Statement}
% The data that supports the findings of this study are available within the article. 
The source code in Julia, which allows for the reproduction of all results in this publication, is publicly available at \href{https://github.com/TheMik1999/Modeling-biases-in-the-generalized-nonlinear-q-voter-model.git}{GitHub}.
% \url{https://github.com/TheMik1999/Modeling-biases-in-the-generalized-nonlinear-q-voter-model.git}
% \bibliography{bibliography}
\bibliography{bib_test}
\end{document}